\documentclass[reprint,aps,pra,amsmath,amssymb,groupedaddress,floatfix,notitlepage]{revtex4-1}
\usepackage{epsfig,psfrag,graphicx,dsfont,mathrsfs,subfigure}
\usepackage[usenames, dvipsnames]{color}
\usepackage[normalem]{ulem}
\newcommand\ket[1]{\left|#1\right\rangle}
 
\newcommand\braket[2]{ \langle #1 | #2 \rangle }
\newcommand\ketbra[2]{ | #1 \rangle\!\langle #2 | }

 \newcommand{\id}{\mathds 1}
\newcommand{\ten}{\otimes} 
\newcommand{\tr}{\mbox{tr}} \renewcommand{\rho}{\varrho}
 
\newcommand{\dEoa}{Q_{A}}
\newcommand{\I}{i}
\newcommand{\blochvol}{V_B}
\newcommand{\hi}{\mathcal{H}}

\begin{document}

\title{Geometric Characterization of True Quantum Decoherence}

\author{Julius Kayser} \author{Kimmo Luoma} \author{Walter T. Strunz} 

\affiliation{Institut f\"{u}r Theoretische Physik, Technische
  Universit\"{a}t Dresden, 01062 Dresden, Germany} 

\date{\today}

\begin{abstract}
  Surprisingly often decoherence is due to classical fluctuations of
  ambient fields and may thus be described in terms of random unitary
  (RU) dynamics. However, there are decoherence channels where such a
  representation cannot exist. Based on a simple and intuitive
  geometric measure for the distance of an extremal channel to the
  convex set of RU channels we are able to characterize the set of
  {\it true quantum} phase-damping channels. Remarkably, using the
  Caley-Menger determinant, our measure may be assessed directly from
  the matrix representation of the channel. We find that the channel
  of maximum quantumness is closely related to a symmetric,
  informationally-complete positive operator-valued measure (SIC-POVM)
  on the environment. Our findings are in line with numerical results
  based on the entanglement of assistance.
\end{abstract}
\maketitle

\section{Introduction}\label{sec:introduction}
In quantum mechanics, any two distinct states of a quantum system may
be coherently superposed to yield a novel state. In addition to the
occupation probabilities of the individual states, such a
superposition is characterized by the coherences which describe its
interference potential. The theory of quantum information processing
exploits the superposition principle in the design of algorithms whose
efficiencies by far exceed conventional, classical schemes
\cite{NielsenChuang2007}. In practice, however, the coherences are
susceptible to decay---a process usually termed decoherence
\cite{Giulini, ZurekRev, Strunz2002}. Sometimes it is possible to
single out a certain basis of robust states which is insensitive to
the given decoherence dynamics (often a simple matter of time scale
\cite{Whaley1998,Braun2001}). With respect to this preferred basis all
populations remain unchanged and only the coherences are subject to
decay. One then also speaks of pure decoherence or phase damping
(a.k.a. dephasing).

In experiment, phase damping is often due to classical fluctuations of
ambient fields (sometimes also called ``random external fields''
\cite{NielsenChuang2007, AlickiLendi1987, LoFranco2015}). These fluctuations have
for example been identified as the main source of decoherence in ion
trap quantum computers, where instabilities are present both in the
trapping fields and in the laser addressing the individual ions
\cite{Monz2009, Monz2010}. Formally, decoherence is then of random
unitary (RU) nature, i.e., it may be represented in terms of a
stochastic ensemble of unitary dynamics. Despite the practical
relevance of this classical, ensemble-based approach, the common
understanding of decoherence is based on the language of open quantum
systems. Here, the loss of coherences is a direct consequence of
growing correlations between the system and its quantum
environment. Note that while these correlations may well turn out to
involve quantum entanglement, surprisingly often the system may have
decohered completely while still being separable from the environment
\cite{EisertPlenio2002, PerniceStrunz2011}.

While RU dynamics may certainly account for a wide range of
experimentally observed phase damping, there are examples of
\emph{true quantum} decoherence, where such a RU representation is
impossible (i.e., the decoherence cannot be attributed to classical
fluctuations) \cite{LandauStreater1993, Buscemi2005,
  HelmStrunz2009}. 
{ The existence of such quantum channels has sparked the 
research in quantum information processing community on the  
asymptotic version of quantum Birkhoff's theorem~\cite{OpenProb} which was
recently solved~\cite{Haagegrup2011}.}
Despite some mentionable effort there is no known
simple criterion allowing to decide whether or not a given dynamics is
RU \cite{Buscemi2006, Audenaert2008, Mendl2009,Xu2012}. The purpose of the
present article is to provide a characterization of the set of true
quantum phase-damping dynamics, i.e., dynamics which may not be
explained in terms of fluctuating fields. Our analysis is based on a
simple and intuitive geometric measure---a volume---for the distance
of a given decoherence dynamics to the convex set of RU dynamics. We
find strong numerical evidence that our results are also valid for
another measure of quantumness based on the entanglement of
assistance.

\section{Preliminaries}\label{sec:preliminaries}
{ In this work we work with finite dimensional Hilbert spaces,
namely $\hi=\mathbb{C}_N$.}
The quantum channel formalism provides a
viable tool to account for a wide variety of open quantum system
dynamics \cite{NielsenChuang2007}. It relies on completely positive,
trace-preserving maps, mapping the initial state of the system 
of
interest onto its time-evolved image, $\mathcal E: \rho \mapsto
\rho'$. Any such map may be written using the Kraus representation
\begin{eqnarray}
  \label{eq:kraus}
  \rho' = \sum_{i} K_i \rho K_i^\dagger, 
\end{eqnarray}
where the trace-preserving character implies $\sum_i K_i^\dagger K_i =
\mathds 1$. With $r$ we denote the Kraus rank of a channel, giving the
number of Kraus operators in (\ref{eq:kraus}) with $\{ K_i \}$ linear
independent. A channel is called {\it unital} if it leaves the
completely mixed state unchanged. Then, the Kraus operators
additionally obey $\sum_i K_i K_i^\dagger = \mathds 1$
\cite{Bengtsson2006}. { We further assume that 
the input and output state spaces of the quantum channel $\mathcal E$ 
have the same dimensionality.}

A prime example is given by a RU channel, where
the $K_i$ may be chosen to be unitary up to a trivial (positive)
factor, that is,
\begin{eqnarray*}
  \rho' = \sum_i p_i U_i \rho U_i^\dagger \quad \left( p_i \geq 0, \sum_i p_i = 1 \right).
\end{eqnarray*}
In addition to the already mentioned experimental significance, RU
channels are frequently studied due to their analytical accessability
\cite{HelmStrunz2010, Alberpapers, Darek2015}.

The case of pure decoherence stands out due to the existence of a
basis $\ket{n}$ of robust states, suggesting the channel may be
written in the form
\begin{eqnarray}
  \label{eq:phase-damping-channel}
  \rho'_{mn} = \braket{a_n}{a_m} \rho_{mn}, 
\end{eqnarray}
with $\{ \ket{a_n} \} \subset \mathds C^r$ being a set of normalized
complex vectors---the {\it dynamical vectors}. It is then sometimes
convenient to introduce the matrix $D$ with $D_{mn} =
\braket{a_n}{a_m}$, such that the channel may be expressed in terms of
the Hadamard product, $\rho' = D \star \rho$, denoting the entry-wise
product of matrices of the same size \cite{HavelHadamard}.

\subsection{True Quantum Phase-Damping}\label{sec:true-quantum-phase}
It is known that the convex set
of unital channels contains non-unitary, extremal channels
\cite{LandauStreater1993, Buscemi2005}. These channels undeniably
represent examples of {\it true} quantum decoherence, i.e.,
decoherence which may not be understood in terms of RU dynamics. A
phase-damping channel is known to be extremal if the projectors onto
the dynamical vectors, $\Pi_n = \ketbra{a_n}{a_n}$, form a (possibly
over-complete) operator basis on the set of $r \times r$ matrices
$\mathcal M_r$ \cite{LandauStreater1993}. This extremality criterion
has a very intuitive, geometric interpretation in terms of the Bloch
representation. Recall that the Bloch vector $\vec b_n \in \mathds
R^{r^2-1}$ corresponding to projector $\Pi_n$ may be assigned via
\begin{eqnarray*}
  \Pi_n = \frac{1}{2} \left( \frac{2}{r} \mathds 1_r 
    + \vec b_n \cdot \vec \sigma \right), 
\end{eqnarray*}
where $\vec \sigma = ( \sigma_1, \ldots, \sigma_{r^2-1})$ is the
vector of a set of orthogonal traceless generators of the SU($r$) with
$\mbox{tr}\sigma_i \sigma_j = 2 \delta_{ij}$ \cite{Bengtsson2006}. A
simple argument shows that extremality is given as soon as the Bloch
vectors span a non-zero volume in $\mathds R^{r^2-1}$
\cite{HelmStrunz2009}. While $r>1$ is certainly necessary for the
channel to be non-unitary, extremality requires $r^2 \leq N$. The
smallest possible dimension allowing for a non-RU channel is thus
given by $N=4$ (a system of two qubits).

\subsection{Measures of Quantumness}\label{sec:measures-quantumness}

Besides plain identification of the
non-classical nature we are also interested in a more quantitative
estimation of the \emph{quantumness} of a given channel (i.e., its
distance to the set of RU channels). The first approach (proposed in
\cite{Audenaert2008}) is based on the entanglement of assistance
\begin{eqnarray*}
  E_A (\rho) &:=& \max_{\{ p_i, \ket{\psi_i} \} }
  \left \{ \sum_i p_i E(\psi_i) \; : \; 
    \sum_i p_i \ketbra{\psi_i}{\psi_i} = \rho \right\}. 
\end{eqnarray*}
It gives the maximum average entanglement entropy $E$ over all
pure-state decompositions of a given state. Recall that via the
Jamiolkowski isomorphism~\cite{HeinosaariZiman} we can assign to every channel $\mathcal E$ a
state $\rho_{\mathcal E}$ by applying $\mathcal E$ to one half of the
maximally entangled state $\sum_n \ket{nn} / \sqrt N$:
\begin{eqnarray*}
  \rho_{\mathcal E} &=& \frac{1}{N} (\mathcal E \otimes \id) \sum_{m,n} \ketbra{mm}{nn}. 
\end{eqnarray*}
Since entanglement is invulnerable to local unitary transformations,
the resulting state is a convex mixture of maximally entangled pure
states if and only if the channel is RU. Trivially, the state's
entanglement of assistance then yields the maximal amount of $ E_A
(\rho_{\mathcal E}) = \log_2 N$, while any smaller value clearly
indicates that the channel is of true quantum type. As a minor
adaption we choose to additionally normalize the result, defining the
``quantumness of assistance'' as
\begin{eqnarray*}
  Q_A (\mathcal E)  = 1 - \frac{E_A (\rho_{\mathcal E})}{\log_2 N}. 
\end{eqnarray*}
Now we have that $ 0 \leq Q_A \leq 1$ and $Q_A = 0$ if and only if the
channel is RU.

As previously discussed, the Bloch representation offers a geometric
approach to identify extremality.  Our previous 
results furthermore suggest
that the volume spanned by the relative states gives a good estimation
of the quantumness of the channel \cite{HelmStrunz2009}. A very
elegant way to directly assess the Bloch volume may be obtained using
the Caley-Menger determinant (see
{{Appendix}}~\ref{sec:assess-bloch-volume}). For 
simplicity let us first
consider the case $r^2 = N$. Then, the $N$ Bloch vectors form a $N-1$
simplex in $\mathds R^{N-1}$ spanning a volume given by
\begin{eqnarray*}
  \blochvol^2 = \frac{(-1)^N}{2^{N-1} ((N-1)!)^2} 
  \left|
    \begin{array}{cccc}
      0 & 1 & \cdots & 1 \\
      1 &      &        &   \\
      \vdots   &  & \! \! \! \! 4 \, ( \mbox{id} - D\star D^*) \!\!\!\!&  \\
      1 &    &  &        
    \end{array}
  \right|. 
\end{eqnarray*}
In the case $r^2 < N$, the direct calculation of the determinant will
yield zero volume at all times. Rather, one has to check the
corresponding determinant for all $r^2$-dimensional submatrices of $D$
obtained by discarding $N-r^2$ rows and columns of the same indices 
{ because extremal channels with rank $r^2<N$ are directly 
related to extremal channels with $r^2=\tilde N <N$.}

\section{Geometric Characterization}\label{sec:geom-char}
The geometric character of the
Bloch volume may now be exploited in order to characterize the set of
two-qubit phase-damping channels. As an additional parameter in our
investigation we consider the purity, which for a quantum state is
defined as the trace of the squared density matrix, $P (\rho ) := \tr
\left\{ \rho^2 \right\}$. In the case of a quantum channel $D$, we may
evaluate the purity of the corresponding Jamiolkowski state,
$P(\rho_D)$. Note that it is straightforward { to see} that a 
unital channel
represents unitary dynamics if and only if $P ( \rho_D ) = 1$. On the
other hand, the minimum value of $P(\rho_D) = 1/N$ is only possible if
$D_{mn} = 0$ for all $m,n=1,\ldots,N$ with $m \ne n$. Then, however,
{\it any} initial state is mapped onto a state with diagonal density
matrix. We may call this channel the {\it completely decohering
  channel}, $D_{cd}$. In this vein, the purity of a channel may thus
be used as an indication of the effect of the channel on any arbitrary
initial state. If it is close to 1, the effect of the channel on the
purity of an arbitrary state is probably small. If it approaches a
value of $1/N$, however, it's impact in terms of purity is possibly
large.

From geometric considerations it is quite trivial to arrive at the
phase-damping channel of maximum Bloch volume: the corresponding Bloch
vectors span a regular tetrahedron inside the Bloch sphere. A possible
choice is given by the four vectors
\begin{eqnarray}
  \label{eq:sic-povm}
  \begin{array}{cc}
    \vec b_1 = \begin{pmatrix} 0 \\ 0 \\ 1 \end{pmatrix},&
    \vec b_2 = \begin{pmatrix} \sin \alpha \\ 0 \\ \cos \alpha \end{pmatrix},\\
    $\,$&$\,$ \\
    \vec b_3 = \begin{pmatrix} \sin \alpha \cos \frac{2 \pi}{3} \\ 
      \sin \alpha \sin \frac{2 \pi}{3} \\ \cos \alpha \end{pmatrix}, \quad&
    \vec b_4 = \begin{pmatrix} \sin \alpha \cos \frac{2 \pi}{3} \\ 
      - \sin \alpha \sin \frac{2 \pi}{3} \\ \cos \alpha \end{pmatrix}, 
  \end{array}
\end{eqnarray}
where $\alpha = \arccos\left( -\frac{1}{3}\right)$ denotes the
so-called tetrahedral angle. With $x:= \sqrt{\frac{ 1+ \cos
    \alpha}{2}} = \sqrt{\frac{1}{3}} \approx 0.57735$, the
corresponding phase damping channel has the matrix representation
\begin{eqnarray}
  \label{eq:Dtetra}
  D_{\Delta} = 
  \begin{pmatrix}
    1 & x & x & x \\
    x & 1 & \I x & -\I x \\
    x & -\I x & 1 & \I x \\
    x & \I x & -\I x & 1
  \end{pmatrix}. 
\end{eqnarray}
Note that this symmetric placement of the Bloch vectors is known from
the concept of SIC-POVMs. In $\mathds C^N$, a SIC-POVM is defined as a
set of $N^2$ normalized vectors $\ket{\psi_i}$ with
$|\braket{\psi_i}{\psi_j}]^2 = \frac{1}{N+1}, \quad i \ne j$
\cite{Renes2004}.

\begin{figure}[t]
  \center
  \includegraphics[width=\linewidth]{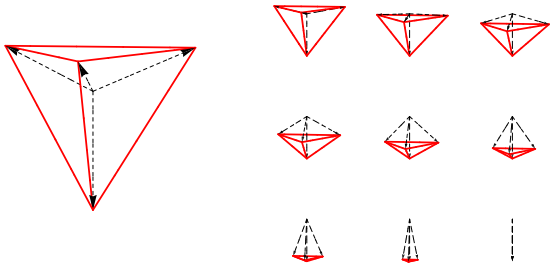}
  \caption{(Color online) Schematic Plot of the Bloch vectors (black,
    dashed arrows) corresponding to the ``maximum quantum'' channel
    (big) and to a selection of MCMQ (small) spanning tetrahedra (red,
    solid lines) of decreasing volume. }
  \label{fig:figure1}
\end{figure}

As previously discussed, a unitary channel has purity $P=1$. The
channel with maximum volume, Eq.~(\ref{eq:Dtetra}), on the other hand,
has purity $P(\rho_{D_\Delta}) = 1/2$. What can be said about channels
with intermediate values of purity? Certainly, there must exist
channels maximizing the volume for a given purity. When expressed in
terms of the Bloch vectors of the dynamical vectors, the Purity of the
channel $D$ is given by $P ( \rho_D ) = \frac{1}{2} \big ( 1 + |\vec
b_S|^2 \big )$, where $\vec b_S = (\vec b_1 + \ldots + \vec b_4)/4$
denotes the barycentre of the Bloch vectors. Of all tetrahedra whose
barycenters lie equidistant from the origin we thus aim to identify
the one spanning the maximum volume. We find this maximum numerically:
it is attained for the Bloch vectors in Eq.~(\ref{eq:sic-povm}), yet
with $\alpha \in [0, \arccos ( -1/3 )]$. For $\alpha = 0$ the volume
is zero, the purity equals one, and the corresponding channel is
unitary.  For $\alpha = \arccos( -1/3 )$ we get the channel with
maximum volume. The values in between give the {\it mixed channels
  with maximum quantumness} (MCMQ). The picture that comes to mind is
the process of closing an umbrella: while one of the Bloch vectors
remains fixed, the three remaining ones move towards the first just
like the metal frame carrying the fabric (Fig.~\ref{fig:figure1}).

\begin{figure}[b]
  \center
  \includegraphics[width=0.8\linewidth]{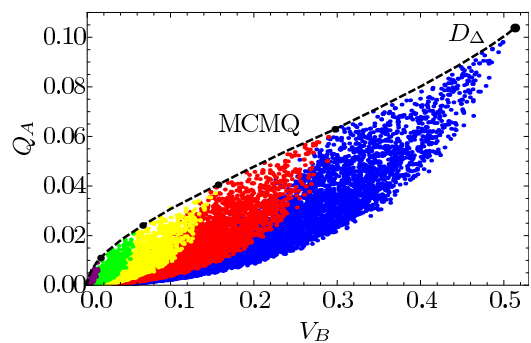}
  \caption{(Color online) Quantumness $\dEoa$ vs. Bloch volume
    $\blochvol$ for a set of random phase-damping channels of rank
    2. The color coding indicates the purity of the channel in the
    following way: blue ($0.5 \leq P < 0.6$), red ($0.6 \leq P <
    0.7$), yellow ($0.7 \leq P < 0.8$), green ($0.8 \leq P < 0.9$),
    purple ($0.9 \leq P < 1$).  The dashed line corresponds to the
    one-parameter set of MCMQ, the channel with maximum Bloch volume
    $D_\Delta$ is indicated with a black dot, as are the MCMQ channels
    with $P=0.6$, $0.7$, $0.8$, and $0.9$ (from right to left).}
  \label{fig:figure2}
\end{figure}

\section{Comparison with the Quantumness of Assistance}
In order to
check the validity of our findings with respect to the quantumness of
assistance we first compare the two measures directly. For this, we
study a set of randomly generated two-qubit phase-damping
channels. The random generation is based on representation
(\ref{eq:phase-damping-channel}): we draw a set of vectors $\ket{a_n},
n = 1, \ldots, 4,$ which are equally distributed on the unit sphere in
$\mathds C^r$. The channel is then obtained as the Gram matrix $\left(
  D_{mn} \right) = \left( \braket{a_m}{a_n} \right)$. At first, only
channels or rank $r=2$ are studied (allowing for the channels to be
extremal in the set of unital channels). To each channel we calculate
Bloch volume $\blochvol$ and quantumness $\dEoa$, as well as the
purity $P$.

We observe that, depending on the purity of the channel, there exist
certain bounds to the quantumness as well as to the Bloch volume: the
lower the purity of the channel, the higher the accessible quantumness
and volume. In Fig.~\ref{fig:figure2} this is highlighted using a
color scheme to single out specific purity intervals. Certainly, there
is no one-to-one correspondence between the two measures; yet, the
correlation is evident. It is quite apparent that towards the channel
of maximum Bloch volume the two measures converge. This is also true
for any of the purity intervals specified. It is thus maybe not too
surprising that the values obtained for the class of MCMQ represents
an upper bound to the accessible quantumness of assistance in terms of
fixed Bloch volume (dashed line in Fig.~\ref{fig:figure2}). These
findings strongly suggest that the MCMQ also maximize the quantumness
of assistance.

\begin{figure}[t]
  \center
  \includegraphics[width=0.8\linewidth]{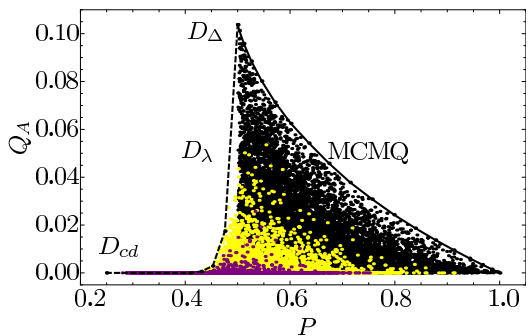}
  \caption{(Color online) Quantumness in terms of Purity for a set of
    random phase-damping channels. The color coding indicates the rank
    of the respective channels: rank 2 (black), rank 3 (yellow), and
    rank 4 (purple). For further details see text. }
  \label{fig:figure3}
\end{figure}

Our findings are best visualized in the Quantumness-Purity-Plane
(Fig.~\ref{fig:figure3}). Here, also channels of rank $3$ and $4$ are
included. For a total number of $70000$ random channels we find no
single violation of the upper bound represented by the MCMQ (solid
line). In addition, we observe that below a purity of $0.5$ there are
only few channels with considerable quantumness. In order to get a
feeling for this behavior we look at channels that are defined as
convex mixture of the channel with maximum volume, $D_{\Delta}$, and
the completely decohering channel, $D_{cd}$: $D_\lambda = (1- \lambda)
D_\Delta + \lambda D_{cd}$. Note that it is easy to see that $D_{cd}$
belongs to the set of RU channels: its RU decomposition is given by $
(\id \ten \id)/4, (\sigma_z \ten \id)/4, (\id \ten \sigma_z)/4,
(\sigma_z \ten \sigma_z)/4$. For this one-parameter class of channels
we numerically estimate the quantumness. We find that the
corresponding quantumness rapidly decays to zero for increasing
$\lambda$ (see the dashed line in Fig.~\ref{fig:figure3}).
 { All sampled channels are RU already for $\lambda > 0.2$. 
This value is considerably lower than the bound $\lambda^*=\frac{14}{15}$
which makes \emph{any} channel RU~\cite{Watrous2008}.}

\section{Summary}\label{sec:summary}
Based on a simple and intuitive geometric
measure---a volume---we are able to characterize the set of true
quantum phase-damping channels acting on two qubits. We identify the
channel with the maximum distance to the set of RU channels, which is
directly linked to the concept of SIC-POVM. In this context our
results imply that the maximally non-classical phase-damping channel
corresponds to a set of rank-one measurements which is best equipped
for distinguishing quantum states on the environment---a point
certainly worth further study. Our findings are in remarkable
agreement with numerical results based on the entanglement of
assistance of the Jamiolkowski state of the channel.

\begin{acknowledgments}
  J.~H. acknowledges financial support by the International Max Planck
  Research School (IMPRS) Dresden.
\end{acknowledgments}
\appendix
\section{Assessing the Bloch volume with the Caley-Menger determinant}
\label{sec:assess-bloch-volume}
The volume of a $N-1$ simplex spanned by the vectors $\vec x_n =
(x_n^{(1)}, x_n^{(2)}, \ldots, x_n^{(N-1)}) \in \mathds R^{N-1}, n =
1, \ldots, N$, may be evaluated using the so-called Caley-Menger
determinant \cite{Blumenthal1970}. With $s_{mn} = \sqrt{ (\vec x_m -
  \vec x_n) \cdot (\vec x_m - \vec x_n)}$ denoting the distance
between vertex $m$ and $n$ it is defined as
\begin{eqnarray}
  \det (A_N) = 
  \left|
    \begin{array}{ccccc}
      0     & 1     & 1     & \cdots & 1     \\
      1     & 0     & s_{1,2}^2 & \cdots & s_{1,N}^2 \\
      1     & s_{1,2}^2 & 0     &  \ddots & \vdots \\
      \vdots & \vdots & \ddots & \ddots & s_{N-1,N}^2 \\
      1    &  s_{1,N}^2 & \cdots & s_{N-1,N}^2 & 0 
    \end{array}
  \right|. 
\end{eqnarray}
The volume of the simplex is then given by
\begin{eqnarray}
  \mbox{Vol}^2 = \frac{(-1)^N}{2^{N-1} ((N-1)!)^2} \det (A_N). 
\end{eqnarray}

A phase-damping channel $D$ in dimension $N$ is defined via $D_{mn} =
\braket{a_n}{a_m}, \quad m,n = 1,\ldots, N$. For a channel of rank $r$
we may assign to each dynamical vector $\ket{a_n} \in \mathds C^r$ a
Bloch vector $\vec b_n \in \mathds R^{r^2-1}$. The mutual distance
between any of these Bloch vectors equates to $s_{m,n}^2 = 4 \left(
  1-\frac{1}{r} \right) - 2 \, \vec b_m \cdot \vec b_n$, while the
matrix elements $D_{mn}$ may be expressed as $|D_{mn}|^2 = |
\braket{a_n}{a_m} |^2 =\frac{1}{r} + \frac{1}{2} \vec b_n \cdot \vec
b_m$.  Put together, it is straightforward to see that $ s_{m,n}^2 = 4
\, \left(1 - |D_{mn}|^2\right),$ so that we arrive at the equivalence
\begin{eqnarray}
  \blochvol^2 = \frac{(-1)^N}{2^{N-1} ((N-1)!)^2} 
  \left|
    \begin{array}{cccc}
      0 & 1 & \cdots & 1 \\
      1 &      &        &   \\
      \vdots   &  & \! \! \! \! 4 \, ( \mbox{id} - D\star D^*) \!\!\!\!&  \\
      1 &    &  &        
    \end{array}
  \right|. 
\end{eqnarray}

\bibliography{mybib}{}

\end{document}